\documentstyle[12pt]{article}

\catcode`\@=11
\long\def\@makefntext#1{
\protect\noindent \hbox to 3.2pt {\hskip-.9pt
$^{{\ninerm\@thefnmark}}$\hfil}#1\hfill}		

\def\@makefnmark{\hbox to 0pt{$^{\@thefnmark}$\hss}}  

\def\ps@myheadings{\let\@mkboth\@gobbletwo
\def\@oddhead{\hbox{}
\rightmark\hfil\ninerm\thepage}
\def\@oddfoot{}\def\@evenhead{\ninerm\thepage\hfil
\leftmark\hbox{}}\def\@evenfoot{}
\def\sectionmark##1{}\def\subsectionmark##1{}}

\newcounter{sectionc}\newcounter{subsectionc}\newcounter{subsubsectionc}
\renewcommand{\section}[1] {\vspace*{0.6cm}\addtocounter{sectionc}{1}
\setcounter{subsectionc}{0}\setcounter{subsubsectionc}{0}\noindent
	{\normalsize\bf\thesectionc. #1}\par\vspace*{0.4cm}}
\renewcommand{\subsection}[1] {\vspace*{0.6cm}\addtocounter{subsectionc}{1}
	\setcounter{subsubsectionc}{0}\noindent
	{\normalsize\it\thesectionc.\thesubsectionc. #1}\par\vspace*{0.4cm}}
\renewcommand{\subsubsection}[1]
{\vspace*{0.6cm}\addtocounter{subsubsectionc}{1}
	\noindent {\normalsize\rm\thesectionc.\thesubsectionc.\thesubsubsectionc.
	#1}\par\vspace*{0.4cm}}

\newcounter{appendixc}
\newcounter{subappendixc}[appendixc]
\newcounter{subsubappendixc}[subappendixc]

\renewcommand{\appendix}[1] {\vspace*{0.6cm}
        \refstepcounter{appendixc}
        \setcounter{figure}{0}
        \setcounter{table}{0}
        \setcounter{equation}{0}
        \renewcommand{\thefigure}{\Alph{appendixc}.\arabic{figure}}
        \renewcommand{\thetable}{\Alph{appendixc}.\arabic{table}}
        \renewcommand{\theappendixc}{\Alph{appendixc}}
        \renewcommand{\theequation}{\Alph{appendixc}.\arabic{equation}}
        \noindent{\bf Appendix \theappendixc #1}\par\vspace*{0.4cm}}

\def\abstracts#1{{

\centering{\begin{minipage}{12.2truecm}\footnotesize\baselineskip=12pt\noindent
	\centerline{\footnotesize ABSTRACT}\vspace*{0.3cm}
	\parindent=0pt #1
	\end{minipage}}\par}}

\renewenvironment{thebibliography}[1]
	{\begin{list}{\arabic{enumi}.}
	{\usecounter{enumi}\setlength{\parsep}{0pt}
\setlength{\leftmargin 1.25cm}{\rightmargin 0pt}
	 \setlength{\itemsep}{0pt} \settowidth
	{\labelwidth}{#1.}\sloppy}}{\end{list}}

\newcounter{itemlistc}
\newcounter{romanlistc}
\newcounter{alphlistc}
\newcounter{arabiclistc}

\newcommand{\fcaption}[1]{
        \refstepcounter{figure}
        \setbox\@tempboxa = \hbox{\footnotesize Fig.~\thefigure. #1}
        \ifdim \wd\@tempboxa > 6in
           {\begin{center}
        \parbox{6in}{\footnotesize\baselineskip=12pt Fig.~\thefigure. #1}
            \end{center}}
        \else
             {\begin{center}
             {\footnotesize Fig.~\thefigure. #1}
              \end{center}}
        \fi}

\newcommand{\tcaption}[1]{
        \refstepcounter{table}
        \setbox\@tempboxa = \hbox{\footnotesize Table~\thetable. #1}
        \ifdim \wd\@tempboxa > 6in
           {\begin{center}
        \parbox{6in}{\footnotesize\baselineskip=12pt Table~\thetable. #1}
            \end{center}}
        \else
             {\begin{center}
             {\footnotesize Table~\thetable. #1}
              \end{center}}
        \fi}

\def\@citex[#1]#2{\if@filesw\immediate\write\@auxout
	{\string\citation{#2}}\fi
\def\@citea{}\@cite{\@for\@citeb:=#2\do
	{\@citea\def\@citea{,}\@ifundefined
	{b@\@citeb}{{\bf ?}\@warning
	{Citation `\@citeb' on page \thepage \space undefined}}
	{\csname b@\@citeb\endcsname}}}{#1}}

\newif\if@cghi
\def\cite{\@cghitrue\@ifnextchar [{\@tempswatrue
	\@citex}{\@tempswafalse\@citex[]}}
\def\citelow{\@cghifalse\@ifnextchar [{\@tempswatrue
	\@citex}{\@tempswafalse\@citex[]}}
\def\@cite#1#2{{$\null^{#1}$\if@tempswa\typeout
	{IJCGA warning: optional citation argument
	ignored: `#2'} \fi}}

 1
 1
 1

\font\ninerm=cmr9

\begin{document}

\centerline{\normalsize\bf DOUBLE SCALING LIMITS, AIRY FUNCTIONS}
\baselineskip=16pt
\centerline{\normalsize\bf AND MULTICRITICAL BEHAVIOUR IN O($N$)}
\baselineskip=16pt
\centerline{\normalsize\bf VECTOR SIGMA MODELS}
\baselineskip=15pt

\vspace*{0.6cm}
\centerline{\footnotesize J. MAEDER and W. R\"UHL}
\baselineskip=13pt
\centerline{\footnotesize\it Department of Physics, University of
Kaiserslautern, P.O.Box 3049}
\centerline{\footnotesize\it 67653 Kaiserslautern, Germany}
\centerline{\footnotesize E-mail: ruehl@gypsy.physik.uni-kl.de}
\vspace*{0.3cm}

\vspace*{0.9cm}
\abstracts{O($N$) vector sigma models possessing catastrophes in their action
are studied.
Coupling the limit $N \rightarrow \infty$ with an appropriate scaling behaviour
of the coupling
constants, the partition function develops a singular factor. This is a
generalized Airy function
in the case of spacetime dimension zero and the partition function of a scalar
field theory for
positive spacetime dimension.}

\message{reelletc.tex (Version 1.0): Befehle zur Darstellung |R  |N, Aufruf
z.B. \string\bbbr}
%
%
%
%
%
\font \smallescriptscriptfont = cmr5
\font \smallescriptfont       = cmr5 at 7pt
\font \smalletextfont         = cmr5 at 10pt
\font \tensans                = cmss10
\font \fivesans               = cmss10 at 5pt
\font \sixsans                = cmss10 at 6pt
\font \sevensans              = cmss10 at 7pt
\font \ninesans               = cmss10 at 9pt
\newfam\sansfam
\textfont\sansfam=\tensans\scriptfont\sansfam=\sevensans
\scriptscriptfont\sansfam=\fivesans
\def\sans{\fam\sansfam\tensans}
\def\bbbr{{\rm I\!R}} 
\def\bbbn{{\rm I\!N}} 
\def\bbbE{{\rm I\!E}} 
\def\bbbm{{\rm I\!M}}
\def\bbbh{{\rm I\!H}}
\def\bbbk{{\rm I\!K}}
\def\bbbd{{\rm I\!D}}
\def\bbbp{{\rm I\!P}}
\def\bbbone{{\mathchoice {\rm 1\mskip-4mu l} {\rm 1\mskip-4mu l}
{\rm 1\mskip-4.5mu l} {\rm 1\mskip-5mu l}}}
\def\bbbc{{\mathchoice {\setbox0=\hbox{$\displaystyle\rm C$}\hbox{\hbox
to0pt{\kern0.4\wd0\vrule height0.9\ht0\hss}\box0}}
{\setbox0=\hbox{$\textstyle\rm C$}\hbox{\hbox
to0pt{\kern0.4\wd0\vrule height0.9\ht0\hss}\box0}}
{\setbox0=\hbox{$\scriptstyle\rm C$}\hbox{\hbox
to0pt{\kern0.4\wd0\vrule height0.9\ht0\hss}\box0}}
{\setbox0=\hbox{$\scriptscriptstyle\rm C$}\hbox{\hbox
to0pt{\kern0.4\wd0\vrule height0.9\ht0\hss}\box0}}}}

\def\bbbe{{\mathchoice {\setbox0=\hbox{\smalletextfont e}\hbox{\raise
0.1\ht0\hbox to0pt{\kern0.4\wd0\vrule width0.3pt height0.7\ht0\hss}\box0}}
{\setbox0=\hbox{\smalletextfont e}\hbox{\raise
0.1\ht0\hbox to0pt{\kern0.4\wd0\vrule width0.3pt height0.7\ht0\hss}\box0}}
{\setbox0=\hbox{\smallescriptfont e}\hbox{\raise
0.1\ht0\hbox to0pt{\kern0.5\wd0\vrule width0.2pt height0.7\ht0\hss}\box0}}
{\setbox0=\hbox{\smallescriptscriptfont e}\hbox{\raise
0.1\ht0\hbox to0pt{\kern0.4\wd0\vrule width0.2pt height0.7\ht0\hss}\box0}}}}

\def\bbbq{{\mathchoice {\setbox0=\hbox{$\displaystyle\rm Q$}\hbox{\raise
0.15\ht0\hbox to0pt{\kern0.4\wd0\vrule height0.8\ht0\hss}\box0}}
{\setbox0=\hbox{$\textstyle\rm Q$}\hbox{\raise
0.15\ht0\hbox to0pt{\kern0.4\wd0\vrule height0.8\ht0\hss}\box0}}
{\setbox0=\hbox{$\scriptstyle\rm Q$}\hbox{\raise
0.15\ht0\hbox to0pt{\kern0.4\wd0\vrule height0.7\ht0\hss}\box0}}
{\setbox0=\hbox{$\scriptscriptstyle\rm Q$}\hbox{\raise
0.15\ht0\hbox to0pt{\kern0.4\wd0\vrule height0.7\ht0\hss}\box0}}}}

\def\bbbt{{\mathchoice {\setbox0=\hbox{$\displaystyle\rm
T$}\hbox{\hbox to0pt{\kern0.3\wd0\vrule height0.9\ht0\hss}\box0}}
{\setbox0=\hbox{$\textstyle\rm T$}\hbox{\hbox
to0pt{\kern0.3\wd0\vrule height0.9\ht0\hss}\box0}}
{\setbox0=\hbox{$\scriptstyle\rm T$}\hbox{\hbox
to0pt{\kern0.3\wd0\vrule height0.9\ht0\hss}\box0}}
{\setbox0=\hbox{$\scriptscriptstyle\rm T$}\hbox{\hbox
to0pt{\kern0.3\wd0\vrule height0.9\ht0\hss}\box0}}}}

\def\bbbs{{\mathchoice
{\setbox0=\hbox{$\displaystyle     \rm S$}\hbox{\raise0.5\ht0\hbox
to0pt{\kern0.35\wd0\vrule height0.45\ht0\hss}\hbox
to0pt{\kern0.55\wd0\vrule height0.5\ht0\hss}\box0}}
{\setbox0=\hbox{$\textstyle        \rm S$}\hbox{\raise0.5\ht0\hbox
to0pt{\kern0.35\wd0\vrule height0.45\ht0\hss}\hbox
to0pt{\kern0.55\wd0\vrule height0.5\ht0\hss}\box0}}
{\setbox0=\hbox{$\scriptstyle      \rm S$}\hbox{\raise0.5\ht0\hbox
to0pt{\kern0.35\wd0\vrule height0.45\ht0\hss}\raise0.05\ht0\hbox
to0pt{\kern0.5\wd0\vrule height0.45\ht0\hss}\box0}}
{\setbox0=\hbox{$\scriptscriptstyle\rm S$}\hbox{\raise0.5\ht0\hbox
to0pt{\kern0.4\wd0\vrule height0.45\ht0\hss}\raise0.05\ht0\hbox
to0pt{\kern0.55\wd0\vrule height0.45\ht0\hss}\box0}}}}

\def\bbbz{{\mathchoice {\hbox{$\sans\textstyle Z\kern-0.4em Z$}}
{\hbox{$\sans\textstyle Z\kern-0.4em Z$}}
{\hbox{$\sans\scriptstyle Z\kern-0.3em Z$}}
{\hbox{$\sans\scriptscriptstyle Z\kern-0.2em Z$}}}}
%
%

 \setlength{\baselineskip}{15pt}
 \section{Introduction}

 Matrix models have been understood as representing stochastic triangulated
 surfaces and thus interpreted as quantum gravity theories. They are treated
 in the ``double scaling limit'' $N \rightarrow \infty,\: g \rightarrow g_c$
 \cite{1}. The same kind of limits can be applied to O($N$) vector models
 \cite{2,3,4} which are connected with statistical ensembles of polymers. These
 models have been studied with the usual $\frac{1}{N}$ expansion and
renormalization
 group techniques \cite{2,3,4,5}. This approach involves a certain amount of
guess
 work.

 If an action possesses a singularity (synonymous: a catastrophe) the double
 scaling limit is naturally defined by letting $N$ tend to infinity and the
 deformation of the singularity go to zero. The partition function develops
then
 a singular factor which is a generalized Airy function if spacetime has
dimension
 zero or a partition function of a scalar field theory if this dimension is
nonzero
 (positive). The Airy function depends in general on $\nu$ scale invariant
parameters,
 where $\nu$ is the codimension of the singularity.

 Singularities can be classified \cite{6} and form $s$-dimensional families.
The $s$
 parameters of these families are called ``moduli''. If $s=0$ the families are
 discrete and are grouped into $A$, $D$, and $E$ series. They are related with
the corresponding
 Lie algebras by their symmetry. The $A$ series can be realized in one-vector
models,
 $D$ and $E$ series need two O($N$) vectors at least. The class of
quasihomogeneous
 singularities is the biggest class we have studied so far and shows some
hitherto
 unobserved structures.

 By application of diffeomorphisms singularities can be brought to a canonical
 form. These diffeomorphisms ``reparametrize'' the coupling constants. The
canonical
 form contains the full information of the singularity and defines the
universality
 class of the multicritical behaviour. The notation of universality class thus
obtains
 mathematical rigour. But the problem of ill-defined (unstable) partition
functions
 can also be approached by the theory of singularites. Stability is either
obtained
 by using complex contours (as used for the Airy function Bi($\zeta$)) or by an
 appropriate infinitesimal deformation. Moreover, renormalization group
equations
 and beta functions can be derived a posteriori.

 In a zerodimensional sigma model not all field degrees of freedom participate
 in the singularity. The residual degrees behave Gaussian at the saddle point
 and are integrated over. They give rise to the regular part of the partition
 function and possess a standard $\frac{1}{N}$ expansion. In sigma models with
 nonzero spacetime dimension small momenta are also treated as deformation
parameters.
 Large momenta are included in the Gaussian variables. The elimination of the
Gaussian
 variables is a nontrivial procedure. It ought to be correct in the singular
 variables at least to the order of the singularity (the method used by Di
Vecchia
 et. al. in \cite{3,4} is wrong). In section 4
  we describe this elimination in a
 nontrivial case.

 \section{Elementary catastrophes in zerodimensional spacetime}

 We outline here the results of \cite{7}. Let $\Phi$ be an O($N$) vector field
 and $d \Phi$ be the Lebesgue measure on $\bbbr_{N}$. Then we define the
partition
 function
 \begin{equation} \label{1}
 Z_{N}(g) = \int d \Phi \exp [-N g(\Phi)]
 \end{equation}
 where $g(\Phi)$ possesses the expansion
 \begin{equation} \label{2}
 g(\Phi) = \frac{1}{2} \sum_{k=1}^{\infty} \frac{g_k}{k} ({\Phi}^{2})^{k}.
 \end{equation}
 Formal, analytic or finite power series are permitted. After integration of
the
 angles we obtain
 \begin{eqnarray} \label{3}
 Z_{N} &=& \frac{{\pi}^{\frac{N}{2}}}{\Gamma(\frac{N}{2})} \int_{0}^{\infty}
 \frac{dx}{x} e^{\frac{N}{2} f(x)} \\
 \label{4} f(x) &=& \mbox{log } x - \sum_{k=1}^{\infty} \frac{g_k}{k} x^{k}.
 \end{eqnarray}
 We normalize $g_1$ to one.

 If we require
 \begin{equation} \label{5}
 g_k = 0, \: k \geq m+1
 \end{equation}
 we have
 \begin{equation} \label{6}
 f^{(m+1)}(x) = (-1)^{m} \frac{m!}{x^{m+1}} \not= 0
 \end{equation}
 and a catastrophe of the type ${A}_{m}$ can be produced at $x_0$ if
 \begin{equation} \label{7}
 f^{(k)}(x_0) = 0 \mbox{ for all } 1 \leq k \leq m.
 \end{equation}
 This is achieved if the coupling constants and $x_0$ are chosen as the
critical
 values
 \begin{eqnarray} \label{8}
 g_{k}^{c} &=& (-1)^{k-1} {m \choose k} m^{-k} \\
 \label{9} x_{0}^{c} &=& m.
 \end{eqnarray}
 From (\ref{8}) we have $g_{m}^{c} > 0$ only if m is odd. Thus the partition
 function (\ref{1}) at the critical point is ill-defined. However, if instead
of
 (\ref{5}) we have
 \begin{equation} \label{10}
 g_{m+1} = \epsilon > 0, \quad g_{k} = 0, \: k \geq m+2
 \end{equation}
 convergence is guaranteed.

 The critical values (\ref{8}), (\ref{9}) define the canonical form of the
 $A_{m}$ catastrophe. We define deformations by
 \begin{eqnarray} \label{11}
 g_k &=& g_{k}^{c} + \Theta_k \quad (2 \leq k \leq m) \\
 \label{12} x_{0} &=& x_{0}^{c} + \Theta_0
 \end{eqnarray}
 so that instead of (\ref{7}) we have (for $1 \leq k \leq m-1$)
 \begin{equation} \label{13}
 f^{(k)}(x_{0}^{c} + \Theta_0) = -(k-1)! \sum_{l=k}^{m} {l-1 \choose k-1}
 m^{l-k} \Theta_l + {\mbox{O}}_{2}(\Theta).
 \end{equation}
 In fact, translation invariance permits us to submit the deformations
(\ref{11}),
 (\ref{12}) to the constraint
 \begin{equation} \label{14}
 f^{(m)}(x_{0}^{c} + \Theta_{0}) = 0
 \end{equation}
 which fixes $\Theta_0$ in terms of $\{ \Theta_{k} \}_{2}^{m}$.

 Now we scale $x$ as
 \begin{equation} \label{15}
 x - x_0 = \lambda \eta
 \end{equation}
 so that $N \rightarrow \infty$ and $\lambda \rightarrow 0$ combine to render
 the $(m+1)$st order term finite and normalized
 \begin{equation} \label{16}
 \frac{N}{2} \frac{\lambda^{m+1}}{(m+1)!} | f^{(m+1)}(m) | = \frac{1}{m+1}.
 \end{equation}
 This implies
 \begin{equation} \label{17}
 \lambda = m \left( \frac{2}{N} \right)^{\frac{1}{m+1}}.
 \end{equation}
 If each linear combination (\ref{13}) scales as
 \begin{equation} \label{18}
 f^{(k)}(m+ \Theta_0) = a_k \left( \frac{2}{N} \right)^{\sigma_k}
 \end{equation}
 for $N \rightarrow \infty$ then for $1 \leq k \leq m-1$
 \begin{equation} \label{19}
 \lim_{N \to \infty} \frac{N}{2} \frac{\lambda^{k}}{k!}
 f^{(k)}(m+ \Theta_0) = \zeta_k
 \end{equation}
 is finite, provided
 \begin{equation} \label{20}
 \sigma_{k} = 1- \frac{k}{m+1}
 \end{equation}
 and
 \begin{equation} \label{21}
 \zeta_k = \frac{m^{k}}{k!} a_k.
 \end{equation}

 The resulting partition function is to leading order in $\frac{1}{N}$
 \begin{equation} \label{22}
 Z_{N} = \frac{\pi^{\frac{N}{2}}}{\Gamma(\frac{N}{2})} e^{\frac{N}{2} f(m)}
 \left( \frac{2}{N} \right)^{\frac{1}{m+1}}
 \cdot Y(\zeta_1, \zeta_2, \ldots \: , \zeta_{m-1})
 \end{equation}
 where $Y$ is the $A_m$-type generalized Airy function
 \begin{equation} \label{23}
 Y(\zeta_1, \zeta_2, \ldots \: , \zeta_{m-1}) = \int_{C^{(m)}} d\eta \exp
 \left\{ \sum_{k=1}^{m-1} \zeta_k \eta^{k} + (-1)^{m} \frac{\eta^{m+1}}{m+1}
\right\}.
 \end{equation}
 If $m$ is odd, $C^{(m)}$ is the real axis with positive orientation. If $m$ is
 even, $C^{(m)}$ is chosen as a linear combination of complex contours each one
 running from infinity to infinity. The singular part of the free energy is
 \begin{equation} \label{24}
 F_{s}(\zeta) = \mbox{log } Y(\zeta_1, \zeta_2, \ldots \: , \zeta_{m-1}).
 \end{equation}
 It satisfies Airy function differential equations and a renormalization group
 equation
 \begin{equation} \label{25}
 \left\{ N \frac{\partial}{\partial N} - \sum_{k=2}^{m} \beta_{k}(\Theta)
 \frac{\partial}{\partial \Theta_k} \right\} F_{s}(\zeta(N,\Theta)) = 0
 \end{equation}
 where (\ref{13}), (\ref{19}) has been substituted for $\zeta_k$. The $\{
\zeta_k \}$
 are the ``double scale invariant'' variables. The beta functions $\{ \beta_k
\}$
 in (\ref{25}) can be calculated to the first order in the $\{ \Theta_k \}$.

 \section{Quasihomogeneous singularities}
 The singularities of Section 2 involve only one variable. If we want
singularities
 with $r$ variables we have to define a corresponding class of more complex
models.
 We shall see that models with $r$ vector fields have the desired properties.

 Let
 \begin{equation} \label{26}
 g(\Phi_1, \Phi_2, \ldots \:, \Phi_{r}) =
 \frac{1}{2} \left\{ \sum_{1 \leq i \leq r} g_{i}(\Phi_{i}^{2}) + \sum_{1 \leq
i < j \leq r}
 s_{ij}( (\Phi_i \Phi_j)^{2} ) \right\}
 \end{equation}
 and
 \begin{equation} \label{27}
 Z_{N}^{(r)} = \int \prod_{i=1}^{r} d \Phi_i \exp \{ - N g(\Phi_1, \Phi_2,
\ldots \:, \Phi_{r}) \}.
 \end{equation}
 We introduce the shorthands
 \begin{eqnarray} \label{28}
 x_i &=& \Phi_i^{2} \\
 \label{29} e_i &=& \frac{\Phi_i}{\sqrt{x_i}} \\
 \label{30} t_{ij} &=& \frac{(\Phi_{i} \Phi_{j})^{2}}{x_i x_j}.
 \end{eqnarray}
 With the uniform measure $d \Omega$ on $S_{N-1}$ define the angular integrals
 \begin{equation} \label{31}
 F(t) = \int \prod_{i=1}^{r} d \Omega(e_i) \prod_{j<k} \delta((e_j \cdot
e_k)^{2} - t_{jk})
 \end{equation}
 where only those $(j,k)$ are included in the product for which $s_{jk}$ is
nonzero.
 If all $s_{jk}$ are nonzero and $N$ is big enough, $F(t)$ (\ref{31}) contains
the
 factor
 \begin{equation} \label{32}
 {\Gamma^{(r)}(e_1, e_2, \ldots \:, e_{r})}^{\frac{N}{2}}
 \end{equation}
 where $\Gamma^{(r)}$ is the Gram determinant of the arguments listed. It is a
 polynomial of $\{ t_{jk}^{\frac{1}{2}} \}$ of degree $r$. For the partition
function
 (\ref{27}) we obtain
 \begin{eqnarray} \label{33}
 Z_{N}^{(r)} &=& C_{N}^{(r)} \int_{\bbbd} \prod_{j<k} dt_{jk} \, h(t) \nonumber
\\
 &\cdot& \int_{(\bbbr_{+})^{r}} \prod_{i=1}^{r} \frac{d x_i}{x_i} \exp \left[
\frac{N}{2}
 \Psi(x_i, t_{jk}) \right]
 \end{eqnarray}
 where $h$ is independent of $N$, $\bbbd$ takes account of the Riemann sheets
necessary,
 and the phase function (reduced action) $\Psi$ is
 \begin{eqnarray} \label{34}
 \Psi &=& \mbox{log} \left\{ (\prod_{i=1}^{r} x_i) \Gamma^{(r)}(e_1, e_2,
\ldots \:, e_r)
 \right\} \nonumber \\
 &-& \sum_{1 \leq i \leq r} g_i(x_i) - \sum_{1 \leq i < j \leq r} s_{ij}(x_i
x_j t_{ij}).
 \end{eqnarray}
 The singular saddle point has to be found in $\bbbd \times (\bbbr_{+})^{r}$.

 Notation is simplified if we use
 \begin{equation} \label{35}
 t_{ij}=t_{ji}, \quad s_{ij}=s_{ji}, \quad \zeta_{ij}=x_i x_j t_{ij}.
 \end{equation}
 Then the saddle point $S$ has to satisfy
 \begin{eqnarray} \label{36}
 x_i {g_i}'(x_i) + \sum_{j (\not= i)} \zeta_{ij} {s_{ij}}'(\zeta_{ij}) = 1, \:
\mbox{ (for all $i$)} \\
 \label{37} \frac{t_{jk}}{\Gamma^{(r)}} \frac{\partial \Gamma^{(r)}}{\partial
t_{jk}} -
 \zeta_{jk} {s_{jk}}'(\zeta_{jk}) = 0, \: \mbox{ (for all $(j,k), \: j<k$)}.
 \end{eqnarray}

 In the generic case this saddle point is Gaussian but we are interested in the
 case in which the Hessian has nonzero corank which is achieved by an
adjustment
 of coupling constants. With
 \begin{eqnarray}
 - x_i^{2} \frac{\partial^{2} \Psi}{\partial x_i^{2}} &=& 1 + x_{i}^{2}
{g_{i}}''(x_i)
 + \sum_{j (\not= i)} {\zeta_{ij}}^{2} {s_{ij}}''(\zeta_{ij}) \nonumber \\
 \label{38} &:=& U_i \\
 - x_i x_j \frac{\partial^{2} \Psi}{\partial x_i \partial x_j} &=& \zeta_{ij}
{s_{ij}}'(\zeta_{ij})
 + \zeta_{ij}^{2} {s_{ij}}''(\zeta_{ij}) \nonumber \\
 \label{39} &:=& \sigma_{ij} \quad (i \not= j)
 \end{eqnarray}
 we obtain for $i \not= j$
 \begin{equation} \label{40}
 - x_i t_{ij} \frac{\partial^{2} \Psi}{\partial x_i \partial t_{ij}} =
\sigma_{ij}
 \end{equation}
 and for $i,j,k$ pairwise different
 \begin{equation} \label{41}
 - x_i t_{jk} \frac{\partial^{2} \Psi}{\partial x_i \partial t_{jk}} = 0.
 \end{equation}
 Thus a corank $r$ of the Hessian can be achieved if in addition to (\ref{36}),
 (\ref{37})
 \begin{eqnarray} \label{42}
 U_i &=& 0 \quad \mbox{ (all $i$)} \\
 \label{43} \sigma_{jk} &=& 0 \quad \mbox{ (all $(j,k), \: j<k$)}.
 \end{eqnarray}
 The location of the singularity $S$ is at $\{ x_i^{(0)}, t_{jk}^{(0)} \}$ and
we
 use normalized variables
 \begin{eqnarray} \label{44}
 \xi_i &=& \frac{x_i - x_i^{(0)}}{x_i^{(0)}} \\
 \label{45} \tau_{jk} &=& \frac{t_{jk} - t_{jk}^{(0)}}{t_{jk}^{(0)}}.
 \end{eqnarray}
 For $r=2$ we have performed the calculation explicitly \cite{7}.

 After integration of the Gaussian degrees of freedom $\{ \tau_{jk} \}$ we have
 an $r$ variable action
 \begin{equation} \label{46}
 \frac{N}{2} f(\xi)
 \end{equation}
 and the singular partition function
 \begin{equation} \label{47}
 Z_{N,s}^{(r)} = \int d^{r} \xi \, h(\xi) \exp \left[ \frac{N}{2} f(\xi)
\right] .
 \end{equation}
 Assume $f(\xi)$ is a deformation of a critical function
 \begin{equation} \label{48}
 f_{c}(\xi) = f_{\delta}(\xi) + f_{> \delta}(\xi)
 \end{equation}
 where $f_{\delta}$ is quasihomogeneous of degree $\delta$ and type $\alpha$
and
 $f_{> \delta}$ has higher degree than $\delta$. To explain these concepts
consider
 a monomial of $\xi$
 \begin{equation} \label{49}
 \xi^{\vec k} = \xi_1^{k_{1}} \xi_2^{k_{2}} \ldots \: \xi_r^{k_{r}}
 \end{equation}
 and apply the multiplicative group $\bbbr_{+}$ to $\xi$ by
 \begin{equation} \label{50}
 \xi_i \rightarrow \lambda^{\alpha_i} \xi_{i}, \: \alpha_{i} \in Q
 \end{equation}
 so that
 \begin{equation} \label{51}
 \xi^{\vec k} \rightarrow \lambda^{\vec{k} \cdot \vec{\alpha}} \xi^{\vec k}.
 \end{equation}
 Then $\xi^{\vec k}$ is called ``of degree $\delta$ and of type
$\vec{\alpha}$'' if
 \begin{equation} \label{52}
 \vec{k} \cdot \vec{\alpha} = \delta.
 \end{equation}
 A polynomial is quasihomogeneous if each monomial contained in it satisfies
 (\ref{52}).

 Consider as an example the singularity $W_{10}$ ($r=s=2$). Its canonical form
 is \cite{6}
 \begin{equation} \label{53}
 \xi_1^{4} + (a_0 + a_1 \xi_2) \xi_1^{2} \xi_2^{3} + \xi_{2}^{6}
 \end{equation}
 with $a_0, a_1$ as moduli ($a_0 \not= \pm 2$) and the principal
quasihomogeneous
 part
 \begin{equation} \label{54}
 f_{\delta}(\xi) = \xi_{1}^{4} + a_0 \xi_{1}^{2} \xi_{2}^{3} + \xi_{2}^{6}
 \end{equation}
 with
 \begin{equation} \label{55}
 \vec{\alpha} = (\frac{1}{4}, \frac{1}{6}), \: \delta=1
 \end{equation}
 \begin{equation} \label{56}
 f_{> \delta}(\xi) = a_1 \xi_{1}^{2} \xi_{2}^{4}.
 \end{equation}
 We will always adjust $\vec{\alpha}$ so that $\delta=1$.

 Now we consider a deformation of $f_{c}(\xi)$
 \begin{equation} \label{57}
 f(\xi) = \sum_{\delta=0}^{1} f_{\delta}(\xi) + f_{> 1}(\xi)
 \end{equation}
 obtained by a change of coupling constants
 \begin{equation} \label{58}
 \{ \gamma_{n}^{c} \} \rightarrow \{ \gamma_{n}^{c} + \Theta_n = \gamma_n \}
 \end{equation}
 and expansion points
 \begin{equation} \label{59}
 \{ \xi_i =0 \}_{1}^{r} \rightarrow \{ \Delta_{i} \}_{1}^{r}.
 \end{equation}
 We can impose $r$ constraints from translation invariance so that the
 $\{ \Delta_{i} \}_{1}^{r}$ are expressed linearly in terms of the $\{
\Theta_{n} \}$.
 The scaling is done such that the degree $\delta=1$ terms absorb the factor
 $N$
 \begin{equation} \label{60}
 \frac{N}{2} \cdot \lambda = 1.
 \end{equation}
 Each term of lower degree $0< \delta <1$
 \begin{equation} \label{61}
 f_{\delta}(\xi) = \sum_{{\vec{k}} \atop {(\vec{\alpha} \cdot \vec{k} =
\delta)}}
 t_{\delta, \vec{k}} \xi^{\vec{k}}
 \end{equation}
 can be expanded in the $\{ \Theta_{n} \}$
 \begin{equation} \label{62}
 t_{\delta, \vec{k}} = \sum_{n} {\cal N}_{\delta, \vec{k}; n} \Theta_{n}
 + {\mbox{O}}_{2}(\Theta)
 \end{equation}
 if it is not kept equal to zero by a translational invariance constraint. Then
 we perform the double scaling
 \begin{equation} \label{63}
 Z_{\delta, \vec{k}} = \lim_{{N \to \infty} \atop {\forall \Theta_n \rightarrow
0}}
 \left( \frac{N}{2} \right)^{1- \delta} t_{\delta, \vec{k}}.
 \end{equation}
 In the singular partition function appears the generalized Airy function
 \begin{eqnarray} \label{64}
 \Phi[Z] &=& \int_{\tilde{C}} d^{r} \xi \exp [- Z(\xi)] \\
 \label{65} Z(\xi) &=& \sum_{0< \delta \leq 1} \sum_{{\vec{k}} \atop
{(\vec{\alpha} \cdot \vec{k} = \delta)}}
 Z_{\delta, \vec{k}} \xi^{\vec{k}}.
 \end{eqnarray}
 Of course the $\delta=1$ part of $Z(\xi)$ can be normalized to the canonical
 form. The number of arguments of $\Phi$ is thus reduced to the codimension of
the
 singularity. If the sigma model studied is too ``narrow'', some of these
arguments
 may be missing (see $X_9$ singularity in \cite{7}). The contour $\tilde{C}$
 belongs to an equivalence class which runs from infinity to infinity and is
thus
 invariant under translations and dilations (\ref{51}).

 In \cite{8} we derive string equations for these models. The singular
partition
 function serves as a ground state functional for the $W$-algebra and is
invariant
 under a ``triangular subalgebra'' depending only on the type $\vec{\alpha}$.
If the
 singularity is homogeneous and
 \begin{equation} \label{66}
 \vec{\alpha} = \alpha_{0} (\frac{1}{r}, \frac{1}{r}, \ldots \:, \frac{1}{r})
 \end{equation}
 this subalgebra is the inhomogeneous $gl(r)$ ($r$ translations in coupling
constant space
 as invariant subalgebra).

 \section{Nonzero spacetime dimensions}

 In the case of spacetime dimensions $D>0$ we obtain partition functions for
$r$ scalar fields
 instead of Airy functions. We will sketch here only the case of the $A_{m}$
singularities
 $(r=1)$. A partly incomplete, partly wrong discussion of these cases was given
by
 Di Vecchia et. al. in \cite{3,4}.

 Consider the action
 \begin{equation} \label{67}
 S = \int d^{D}x \left\{ \frac{1}{2} \partial_{\mu} \Phi \partial_{\mu} \Phi
(x) + \frac{1}{2}
 \beta^{2} \Phi^{2}(x) + U(\Phi^{2})(x) \right\}
 \end{equation}
 with the potential
 \begin{equation} \label{68}
 U(\sigma) = \sum_{n=2}^{\infty} \frac{f_n}{n} \sigma^{n}.
 \end{equation}
 Then a standard trick gives us the partition function in terms of two scalar
 fields $\sigma$ and $\rho$
 \begin{equation} \label{69}
 Z = \int D\sigma D\rho \exp [ -N S_{eff}(\sigma,\rho)]
 \end{equation}
 with
 \begin{eqnarray}
 S_{eff} &=& \int d^{D}x [ U(\sigma)(x) - i \sigma(x) \rho(x)] \nonumber \\
 \label{70} &+& \frac{1}{2} \mbox{Tr log } [-\Delta + \beta^{2} + 2i \rho].
 \end{eqnarray}
 The saddle point is assumed at constant values $\sigma_{0}, \rho_{0}$
 \begin{eqnarray} \label{71}
 U'(\sigma_0) &=& i \rho_0 \\
 \label{72} \sigma_0 &=& \int \frac{d^{D}p}{(2 \pi)^{D}} (p^{2}+ m^{2})^{-1} \\
 \label{73} m^{2} &=& \beta^{2} + 2i \rho_0.
 \end{eqnarray}
 The dimensions $D$ are interpolated and firstly restricted to the interval
 $0<D<2$, so that renormalizations are not necessary.

 The fields $\sigma$ and $\rho$ fluctuate around their stationary values
(\ref{71})-(\ref{73})
 \begin{eqnarray} \label{74}
 \sigma(x) &=& \sigma_0 (1+\alpha(x)) \\
 \label{75} \rho(x) &=& \rho_0 (1+\beta(x)).
 \end{eqnarray}
 In terms of the Fourier transforms of $\alpha$ and $\beta$ the Hessian is
 \begin{eqnarray}
 S_{eff}^{(2)} &=& \frac{1}{2} \int \frac{d^{D}k}{(2 \pi)^{D}}
(\hat{\alpha}(-k), \hat{\beta}(-k))  \nonumber \\
 \label{76} & & \left( \begin{array}{cc} \sigma_{0}^{2} U''(\sigma_0) & -i
\sigma_0 \rho_0 \\
 -i \sigma_0 \rho_0 & 2 \rho_{0}^{2} \Sigma(k) \end{array} \right)
 \left( \begin{array}{c} \hat{\alpha}(k) \\ \hat{\beta}(k) \end{array} \right)
 \end{eqnarray}
 where
 \begin{equation} \label{77}
 \Sigma(k) = \int \frac{d^{D}p}{(2 \pi)^{D}} [(p^{2}+ m^{2})((p-k)^{2}+
m^{2})]^{-1}.
 \end{equation}
 The Hessian is diagonalized by
 \begin{equation} \label{78}
 \left( \begin{array}{c} \hat{\alpha}(k) \\ \hat{\beta}(k) \end{array} \right)
=
 \left( \begin{array}{c} {a}(k) \\ i \end{array} \right) \hat{\xi}(k) +
 \left( \begin{array}{c} -i {b}(k) \\ 1 \end{array} \right) \hat{\eta}(k)
 \end{equation}
 \begin{eqnarray}
 S_{eff}^{(2)} &=& \frac{1}{2} \int \frac{d^{D}k}{(2 \pi)^{D}} \left\{
\lambda_{+}(k) N_{+}(k) \hat{\xi}(-k) \hat{\xi}(k) \right. \nonumber \\
 \label{79} & & + \left. \lambda_{-}(k) N_{-}(k) \hat{\eta}(-k) \hat{\eta}(k)
\right\}
 \end{eqnarray}
 where $N_{\pm}(k)$ are the squares of the eigenvectors
 \begin{equation} \label{80}
 N_{+}(k) = {a(k)}^{2}-1, \quad N_{-}(k) = 1-{b(k)}^{2}
 \end{equation}
 and it turns out that
 \begin{equation} \label{81}
 a(k) b(k) = 1.
 \end{equation}
 The eigenvalues $\lambda_{\pm}(k)$ are
 \begin{eqnarray}
 \lambda_{\pm}(k) &=& \frac{1}{2} \left\{ U''(\sigma_{0}) \sigma_{0}^{2} + 2
\rho_{0}^{2} \Sigma(k) \right. \nonumber \\
 \label{82} &\pm& \left. [(U''(\sigma_{0}) \sigma_{0}^{2} - 2 \rho_{0}^{2}
\Sigma(k))^{2} - 4 \sigma_{0}^{2} \rho_{0}^{2}]^{\frac{1}{2}} \right\}.
 \end{eqnarray}

 We will make $\lambda_{-}(k)$ zero at $k=0$. Then $\lambda_{+}(k)$ is positive
and
 $\hat{\xi}(k)$ a Gaussian degree of freedom. If we put the system into a box
with
 periodic boundary conditions, then the spectrum of $k$ is discrete and only
$\hat{\eta}(0)$
 is non-Gaussian, whereas $\hat{\eta}(k), \: k \not= 0$, are Gaussian. In this
case the
 saddle point for the Gaussian integration is determined by
 \begin{eqnarray} \label{83}
 \frac{\partial S_{eff}}{\partial \hat{\xi}(k)} &=& 0, \: \mbox{all } k \\
 \label{84} \frac{\partial S_{eff}}{\partial \hat{\eta}(k)} &=& 0, \: \mbox{all
} k \not= 0
 \end{eqnarray}
 and its location depends on $\hat{\eta}(0)$. Because of translational
invariance these
 equations are solved by
 \begin{equation} \label{85}
 \hat{\xi}(k) = \hat{\eta}(k) = 0 \: \mbox{ for } k \not= 0
 \end{equation}
 and with
 \begin{equation} \label{86}
 \left. S_{eff} \right|_{\hat{\xi}(k)=\hat{\eta}(k)=0, \: k \not= 0} =
 {\tilde{S}}_{red}(\hat{\xi}(0), \hat{\eta}(0))
 \end{equation}
 by additionally
 \begin{equation} \label{87}
 \frac{\partial}{\partial \hat{\xi}(0)} {\tilde{S}}_{red}(\hat{\xi}(0),
\hat{\eta}(0)) = 0
 \end{equation}
 which means
 \begin{equation} \label{88}
 - \lambda_{+}(0) N_{+}(0) \hat{\xi}(0) = \sum_{n=3}^{\infty}
\frac{\partial}{\partial \hat{\xi}(0)}
 {\tilde{S}}_{red}^{n}(\hat{\xi}(0), \hat{\eta}(0)).
 \end{equation}
 Here the superscript $n$ denotes the order in both arguments.

 An iterative solution of (\ref{88}) gives
 \begin{equation} \label{89}
 \hat{\xi}(0) = H(\hat{\eta}(0)) = \sum_{l=2}^{\infty} a_l {\hat{\eta}(0)}^{l}
 \end{equation}
 and
 \begin{eqnarray}
 {{S}}_{red}(\hat{\eta}(0)) &=& {\tilde{S}}_{red}(H(\hat{\eta}(0)),
\hat{\eta}(0)) \nonumber \\
 \label{90} &=& \sum_{n=2}^{\infty} \frac{g_n}{n} {\hat{\eta}(0)}^{n}.
 \end{eqnarray}
 Thus after the Gaussian integration we end up with a zerodimensional case as
treated
 in Section 2. First one calculates the critical coupling constants $g_{n}^{c}$
in
 (\ref{90}) and in turn the critical constants $f_{n}^{c}$ in the potential
$U(\sigma)$.

 Now we turn to the case of physical interest, namely infinite volume and a
continuous
 momentum spectrum. In the case of the singularity $A_m$ in (\ref{90}) the
coupling
 constants are deformed as usual
 \begin{equation} \label{91}
 f_n = f_{n}^{c} + \Theta_n \: (2 \leq n \leq m).
 \end{equation}
 In addition the domain of small momenta
 \begin{equation} \label{92}
 |k| < \Lambda
 \end{equation}
 is considered as deformation, whereas large momenta
 \begin{equation} \label{93}
 |k| > \Lambda
 \end{equation}
 belong to Gaussian degrees of freedom $\hat{\xi}(k), \hat{\eta}(k)$ and are
integrated
 over. Our aim is a partition function of a scalar field $\phi$ in conventional
form
 \begin{eqnarray}
 \int D \phi \exp \left\{ - \left[ \frac{1}{2} \int d^{D}x \, \phi(x)
(-\Delta+M^{2}) \phi(x) \right. \right. \nonumber \\
 \label{94} + \sum_{l=3}^{m} \zeta_{l} \int d^{D}x \, {\phi(x)}^{l} +
\frac{F_{m+1}}{m+1}
 \left. \left. \int d^{D}x \, {\phi(x)}^{m+1} \right] \right\}
 \end{eqnarray}
 which replaces the Airy function (\ref{23}). Contrary to the Airy function
there is
 no linear term in $\phi$, there is a term $\phi^{m}$, and normalization is
applied
 to the kinetic energy and not to $\phi^{m+1}$. Moreover the mass term $M^{2}
\phi^{2}$
 must be positive. This can be achieved by approaching the canonical form of
the
 singularity in the neighbourhood of deformations on a specific path, namely on
a
 surface of stable $A_1$ singularities contingent to $A_m$.

 In the first step we reduce $S_{eff}$ to $S_{red}(\hat{\eta}(k)), \: |k|<
\Lambda$,
 that possesses a power series expansion analogous to (\ref{90}) in terms of
convolution powers
 $(\hat{\eta})_{*}^{n}$ for $n \geq 3$. We expand $\lambda_{-}(k)$ to first
order in $k^{2}$
 and the coupling constant deformations $\{ \Theta_n \}$ (\ref{91}). Then we
scale $k$ by
 \begin{eqnarray} \label{95}
 k &=& N^{-\lambda} k' \\
 \label{96} x &=& N^{\lambda} x'.
 \end{eqnarray}
 The exponent $\lambda$ must be positive to map the domain of small $k$ on
$\bbbr_{D}$
 for large $N$. As in Section 2 wee firstly examine the term $\phi^{m+1}$ which
 entails
 \begin{equation} \label{97}
 \phi(x') = C^{(m)} N^{\frac{1+D \lambda}{m+1}} \eta(x).
 \end{equation}
 With this ansatz we enter the kinetic energy term and find
 \begin{equation} \label{98}
 \lambda = \frac{m-1}{2(m+1)-D(m-1)}
 \end{equation}
 and
 \begin{equation} \label{99}
 C^{(m)} = \sqrt{ 2 \Pi_{2} (i \rho_0)^{2} \frac{1}{6} (2-\frac{D}{2})
\frac{1}{m^{2}} }
 \end{equation}
 with
 \begin{equation} \label{100}
 \Pi_{n} = \int \frac{d^{D}k}{(2 \pi)^{D}} (k^{2} + m^{2})^{-n}.
 \end{equation}
 The mass $M^{2}$ is obtained from a double scaling constraint on the coupling
constants
 \begin{equation} \label{101}
 \lim_{{N \to \infty} \atop {\forall \Theta \rightarrow 0}}
 \left( \sum_{n=2}^{m} (n-1) \Theta_n \sigma_{0}^{n-2} \right)
 = \frac{1}{2 \Pi_{2}} \frac{1}{6} (2- \frac{D}{2}) \frac{M^{2}}{m^{2}}.
 \end{equation}
 This constraint fixes the limiting contour on the stable $A_1$ surface.

 From (\ref{91}) we obtain for the intermediate coupling constant $g_n$ in
 (\ref{90})
 \begin{equation} \label{102}
 g_n - g_{n}^{c} = n \sum_{l=2}^{m} \alpha_{nl}^{(m)} \Theta_{l}.
 \end{equation}
 Then the double scaling limit defines the scale invariant variables $\zeta_n$
 \begin{equation} \label{103}
 \zeta_n = \lim_{{N \to \infty} \atop {\forall \Theta \rightarrow 0}}
(C^{(m)})^{-n} N^{\chi_{n}^{(m)}}
 \sum_{l=2}^{m} \alpha_{nl}^{(m)} \Theta_{l}
 \end{equation}
 where the critical indices are
 \begin{eqnarray} \label{104}
 \chi_{n}^{(m)} &=& \frac{2(m+1-n)}{2(m+1)-D(m-1)} \\
 \label{105} \chi_{2}^{(m)} &=& 2 \lambda \quad \mbox{(see (\ref{98}))}.
 \end{eqnarray}

 \noindent The coupling constant $F_{m+1}$ in (\ref{94}) is a critical value
and thereby fixed
 \begin{equation} \label{106}
 F_{m+1} = g_{m+1}^{c}(C^{(m)})^{-m-1}.
 \end{equation}
 However, this coupling constant varies with the mass of the fundamental field
 $\Phi$. \pagebreak

 Now we consider the interval of dimensions $2<D<4$. Then $\Pi_1$ but no
$\Pi_n, \: n \geq 2$
 (\ref{100}), diverges. This entails that $\sigma_0, \rho_0$ and the bare
coupling
 constants $f_n$ are infinite and must be renormalized. First we continue
$\Pi_1$
 analytically from $0<D<2$ to $2<D<4$ and call this function $\Pi_{1}^{an}$
 \begin{equation} \label{107}
 \Pi_{1}^{an} = \int \frac{d^{D}p}{(2 \pi)^{D}} [(p^{2} + m^{2})^{-1} -
(p^{2})^{-1}].
 \end{equation}
 Then we replace the saddle point equation (\ref{72}) by the renormalized
equation
 \begin{eqnarray} \label{108}
 \sigma_{0}^{ren} &=& \Pi_{1}^{an} = \sigma_{0} - \sigma_{\infty} \\
 \label{109} \sigma_{\infty} &=& \int \frac{d^{D}p}{(2 \pi)^{D}} (p^{2})^{-1}
\: \mbox{(divergent)}.
 \end{eqnarray}
 We insert $\sigma_{0}$ (\ref{108}) into the potential (\ref{68}) and reorder
to
 obtain
 \begin{eqnarray} \label{110}
 U'(\sigma_0) &=& \frac{1}{2} \Delta m^{2} + \sum_{n=2}^{m} f_{n}^{ren}
(\sigma_{0}^{ren})^{n-1} \\
 \label{111} &=& \frac{1}{2} \Delta m^{2} + {U^{ren}}'(\sigma_{0}^{ren}).
 \end{eqnarray}
 The divergent term $\Delta m^{2}$ is absorbed in $i \rho_{0}^{ren}$ and
finally in the
 renormalized mass.

 After these renormalizations all arguments go through as long as $\lambda$
(\ref{98})
 is positive, i.e. for
 \begin{equation} \label{112}
 D < D_{\infty} = 2 \, \frac{m+1}{m-1}.
 \end{equation}
 But this is the classical condition for the partition function (\ref{94}) to
be
 superrenormalizable.

 For the renormalizable case $D=D_{\infty}$ the scaling procedure is not
defined due
 to the divergence of the critical indices $\chi_{n}^{(m)}$ (\ref{104}) and
$\lambda$
 (\ref{98}). We can however renormalize the parameter $N$ by introducing
 \begin{equation} \label{113}
 N'(D) = N^{\frac{1}{2(m+1) - D(m-1)}}
 \end{equation}
 so that the scaling procedure defined in terms of $N'$ is analytic in $D$.
Therefore
 we propose to treat the renormalizable case for all $2 < D < 4$ by analytic
continuation
 in $D$ as well.

 Details about how the results presented in this section are derived and more
details can
 be found in \cite{9}. 

 \end{document}